# Errors in AI-Assisted Retrieval of Medical Literature: A Comparative Study

Jenny Gao[1], Yongfeng Zhang[2], Mary L Disis,[3] Lanjing Zhang[4,5,6] *

[1]College of Arts and Science, New York University, New York, NY 10003

[2] Department of Computer Sciences, School of Arts & Sciences, Rutgers University, Piscataway, NJ, USA.

[3]UW Medicine Cancer Vaccine Institute University of Washington, Seattle, WA 98109, United States.

[4] Department of Chemical Biology, Ernest Mario School of Pharmacy, Rutgers University, Piscataway, NJ, USA.

[5] Department of Pathology, Princeton Medical Center, Plainsboro, NJ, USA.

[6] Rutgers Cancer Institute, New Brunswick, NJ, USA.

* Correspondence: Lanjing Zhang, MD, Department of Chemical Biology, Ernest Mario School of Pharmacy, Rutgers University, 164 Frelinghuysen Road Piscataway, NJ 08854 NJ 08536, USA. Tel: 609-853-6833, Tax: 609-853-6841, Email: lanjing.zhang@rutgers.edu

## Abstract (300 words)

Large language models (LLMs) assisted literature retrieval may lead to erroneous or fabricated references. However, these errors have not been rigorously quantified. Therefore, we quantitatively compare errors in reference retrieval across 5 widely used free-version LLM platforms and identify the factors associated with retrieval errors. We evaluated the 2,000 references retrieved by 5 LLMs (Grok-2, ChatGPT GPT-4.1, Google Gemini Flash 2.5, Perplexity AI, and DeepSeek GPT-4) for 40 randomly-selected original articles (10 per journal) published between January 2024 and July 2025 from *British Medical Journal (BMJ), Journal of the American Medical Association, and The New England Journal of Medicine (NEJM).* Primary outcomes were a multimetric score ratio combining validity of digital object identifier, PubMed ID, Google-Scholar link, and relevance; and complete miss rate (proportion of references failing all applicable metrics). Multivariable regression was used to examine independent associations. LLM platforms completely failed to retrieve correct reference data 47.8% of the time. The average score ratio of the 5 LLM platforms was 0.29 (standard deviation, 0.35; range, 0-1.25), with a higher score ratio indicating a higher accuracy in retrieving relevant references and correct bibliographic data. The highest and lowest accuracies were achieved by Grok (0.57) and Genimi (0.11), respectively. Compared with *BMJ*, *NEJM* articles had lower score ratios and higher complete miss rates (P<.001). In multivariable analysis, both LLM platforms and journals were independently associated with score ratios and complete miss rate, respectively. This quantitative comparative study of LLM-assisted medical literature showed modest overall performance of LLMs and significant variability in retrieval accuracy across platforms and journals. LLM platforms and journals are associated with LLM's performance in retrieving medical literature. Bibliographic data should be carefully reviewed when using LLM-assisted literature retrieval.

## Introduction

The rapid proliferation of large language models (LLMs) has transformed numerous fields, including biomedicine and bioinformatics, where they offer potential for accelerating literature synthesis, reference retrieval, and knowledge discovery.[1, 2] LLMs can assist in literature synthesis tasks such as generating relevant citations for research articles.[3, 4] LLM-based ensemble strategies with retrieval-augmented generation prompts can also help literature screening.[5] However, LLMs often produce plausible but fabricated or inaccurate content, including fictitious references and incorrect digital object identifier (DOI).[6-10] Some scholars refer the observation as referential hallucination.[8, 11-14]

High rates of referential hallucination have been documented across early LLM versions, with studies reporting fabrication in up to 50-80% of generated citations in biomedical contexts.[14, 15] Literature search errors have been observed not only in factual claims but bibliographic metadata, including DOIs and journal citations.[8, 14] This observation raises concerns about reliability in high-stakes applications like medical research, where inaccurate references could propagate errors or undermine scholarly integrity.[16, 17] However, there are few quantitative comparative or association analyses on the error/hallucination rates of contemporary LLMs.

Therefore, we quantified the errors of five widely used free-version LLM platforms (Grok, ChatGPT, Google Gemini, Perplexity AI, and DeepSeek) in searching and retrieving references relevant to original research articles published in four leading medical journals, including *British Medical Journal (BMJ)*, *Journal of the American Medical Association (JAMA)*, *The Lancet* and *The New England Journal of Medicine (NEJM)*. By randomly sampling articles from 2024 and early 2025 and rigorously validating 2,000 retrieved references using a multimetric framework, we aimed to provide an objective and reproducible assessment of errors in LLM-assisted medical literature search and retrieval. Our findings highlight significant variability in LLM performance and provide insights into the factors associated with these errors.

## Materials and Methods
**Randomly select original articles**

Five original research articles published in 2024 and 5 in early 2025 (January 1 to July 15) were randomly selected from each of the four leading medical journals: *BMJ*, *JAMA*, *Lancet* and *NEJM* (**Figure 1**). Specifically, all original articles published in these journals in 2024 and early 2025 were assigned a number sequentially according to their PubMed indexing times. We used the Google's random number generator to generate 5 random numbers for the 2024 cohort and another 5 for the 2025 cohort of each journal. Only the articles corresponding to these random numbers were included in the analysis.

**Retrieve relevant articles using LLM platforms**

For each of the 40 original articles, we prompted 5 commonly used LLM platforms (all free-versions) with the article's abstract to retrieve 10 relevant references from July 15 to August 30, 2025 and later December 20 to December 30, 2025, including Grok (Grok-2), ChatGPT (GPT-4.1), Google Gemini (Flash 2.5), Perplexity AI, and DeepSeek (GPT-4). Each query focused on one article at the time.

The prompt was "Generate a list of 10 key references related to this abstract with the title, year published, DOI, PubMed ID, and Google Scholar URL; no description; clearly indented for each reference" with the article's abstract present at the end of the prompt. We then instructed the LLM platforms to confirm these publication identification data using the prompt "Please confirm the DOI, PubMed ID and Google Scholar link." The results confirmed by the platform were recorded and subject to manual checking.

**Manually validate retrieved references**

Four validity metrics were manually generated in two strategies (**Figure 1**).

We first searched the databases of DOI (doi.org), PubMed (pubmed.gov) and Google Scholar with the retrieved reference's title and authors. If a DOI, PubMed ID or Google Scholar link was resulted and matched to that provided by the LLM platform, one point would be given to the respective metric. If resulted but not matched, no points would be given. If no entry was resulted, the metric would be not applicable. For example, an engineering reference/article understandably might not be included in PubMed database and hence had no PubMed ID. Three of the 4 metrics were assigned using this strategy.

In the other strategy, we searched the title and reference list of the index article with the retrieved references to assign relevance score. If the title and first author of the retrieved reference were found in the reference list, one point of the relevance score would be given. If not found, no points would be given. However, if the retrieved reference was the index article, two points would be given.

**Calculate score ratio and complete miss status**

Because some retrieved references had not-applicable metrics, we used a composite ratio and complete miss status to more objectively evaluate the retrieved references.

First, we generated the total score by totaling the scores of DOI, PubMed ID, Google Scholar link and relevance score for each retrieved reference. Second, we generated the score cap by counting the number of applicable validity metrics. The retrieved references with not-applicable metric(s) then had a smaller score cap than those without. Third, to adjust bonus points in relevance score, we added 1 to the score if the retrieved reference was the index article. Finally, we divided the total score by the score cap to generate the score ratio for each retrieved reference. To highlight the retrieved references without any validated metrics, the complete miss status would be given if the retrieved reference's total score was zero.

**Regression analyses and statistics**

Multivariable regression analyses were performed to identify the factors associated to the score ratio and complete miss, respectively, including only the factors associated with either of them in univariable regression. The factors of interest were journal, publication year and LLM platform.

Chi-square or Fisher exact test was used to compare categorical data by metrics. We performed all statistical analyses and plotting with Stata (version 18. College Station, TX: StataCorp LLC.) and defined statistical significance using $p < 0.05$.

# Results

**Random selection of the original articles**

The 4 leading medical journals published 1,187 original articles from January 1 to December 31, 2024 and 692 from January 1 to July 15, 2025, from which we randomly selected qualified original articles (**Figure 1**).

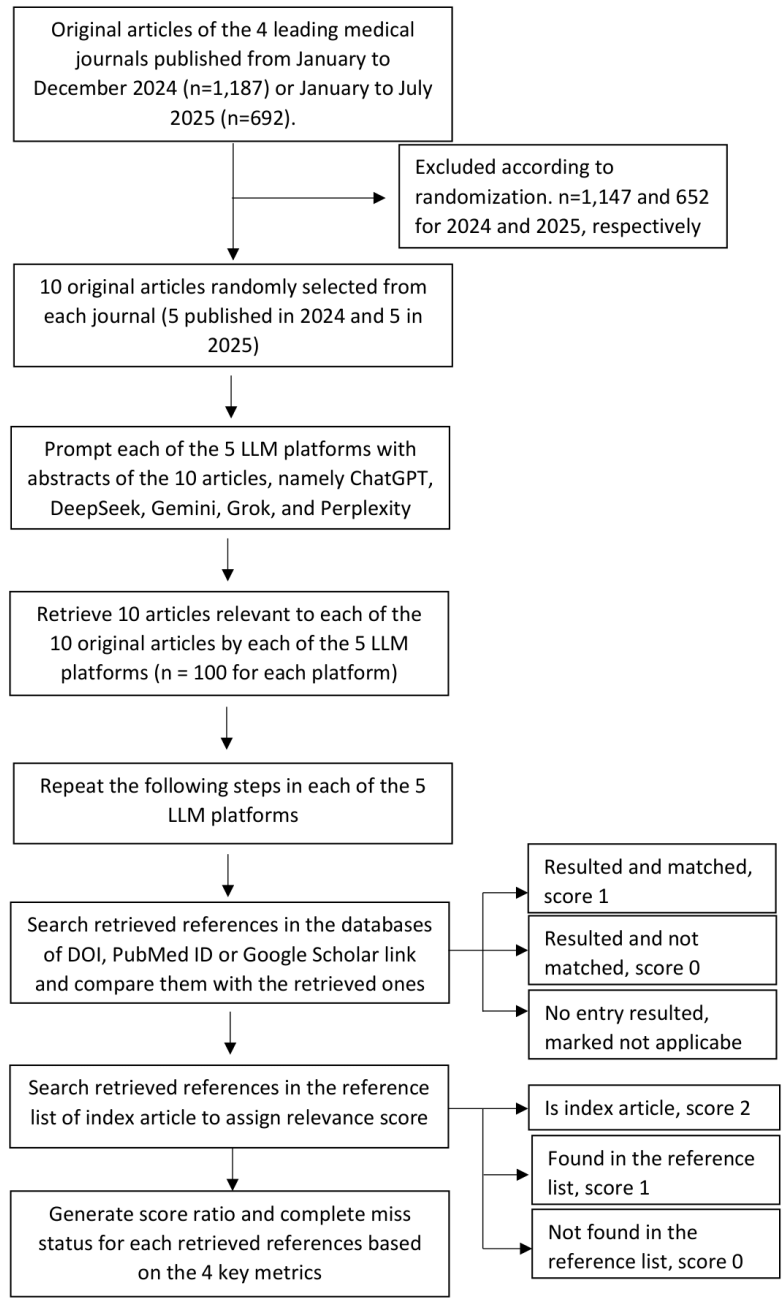

**Figure 1. Flowchart of the study.** The four leading medical journals include *British Medical Journal, Journal of the American Medical Association, Lancet* and *The New England Journal of Medicine,* and. The four key accuracy metrics were digital object identifier, PubMed ID, and Google Scholar link.

**Score ratio**

The average score ratio of the 5 LLM platforms (n=400 for each platform) was 0.29 (standard deviation, 0.35; range, 0-1.25), with a higher score ratio indicating a higher accuracy in retrieving bibliographic data of medical literature. The highest and lowest score ratios were achieved by Grok (0.57) and Genimi (0.11), respectively (**Figure 2**), while the journal *BMJ* (0.33) appeared to have a higher score ratio than *Lancet* (0.29), *NEJM* (0.25) and *JAMA* (0.29).

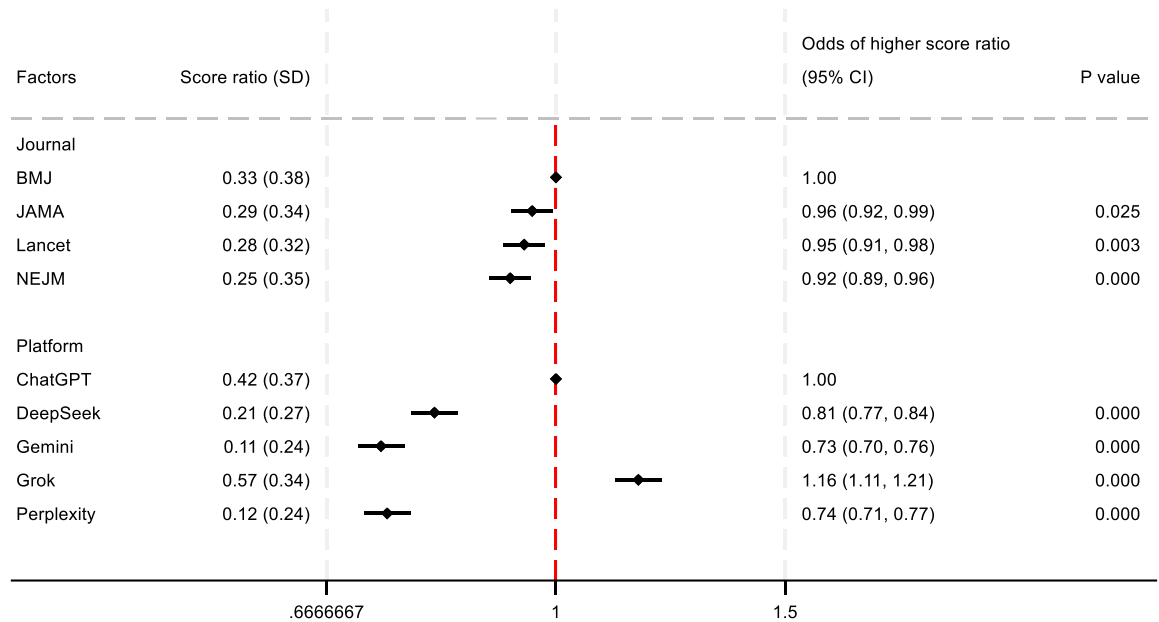

**Figure 2. Independent associations of journal and large-language model platform with score ratio**. The forest plot shows that journal and large-language model (LLM) platform were independently associated with score ratio in multivariable regression analysis. Odds ratios with 95% confidence intervals (CI) were shown. The mean and standard deviation (SD) of the score ratios by journal and LLM platform are listed on the left side, respectively. *BMJ, British Medical Journal; JAMA, Journal of the American Medical Association; NEJM, The New England Journal of Medicine.*

Journal and LLM platform were both independently linked to score ratio in multivariable regression analyses (**Figure 2**), while publication year was not associated with score ratio even in univariable regression. Compared with *BMJ*, *JAMA, Lancet* and *NEJM* were all associated with lower score ratio. Grok was linked to a higher score ratio than ChatGPT (p<0.001), while all other platforms were associated with lower ones (p<0.001 for all).

**Complete miss rate**

LLM platforms completely failed to retrieve a correct reference 47.8% of the time as shown by the mean complete miss rate (**Table 1**). The best complete miss rate was achieved by Grok (11.2%, **Figure 3**), while the worse by Genimi (78.5%). Consistent with the score ratio data, *NEJM* had the highest complete miss rate (56.0%), while others remained similar to each other.

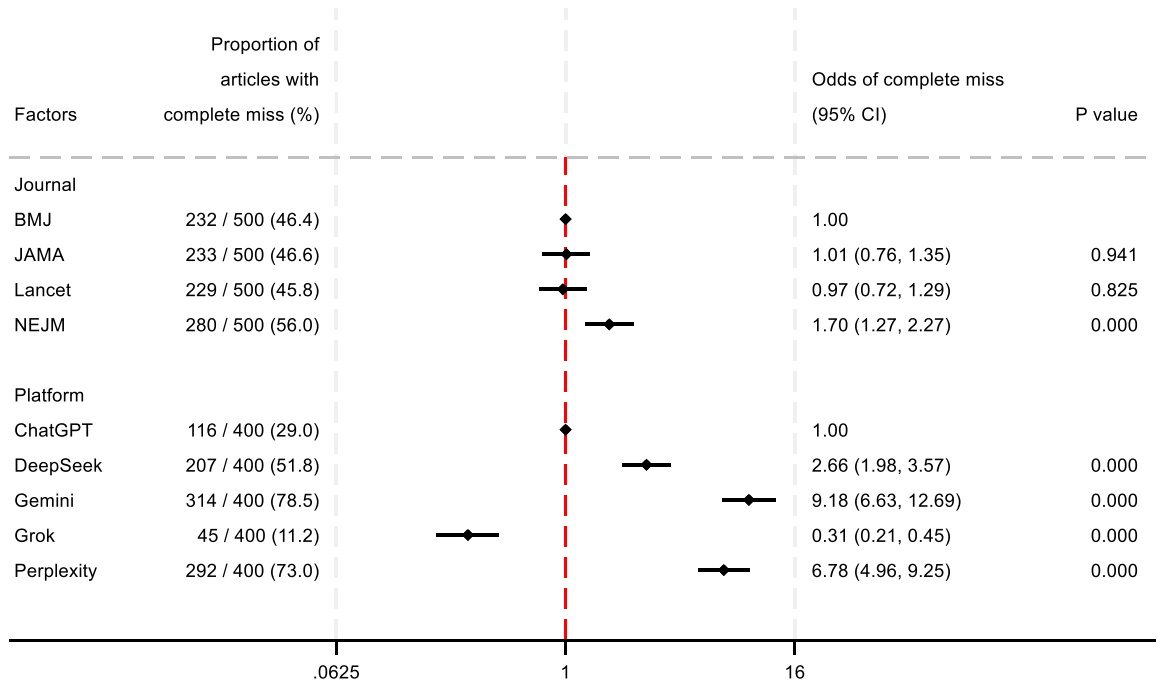

**Figure 3. Independent associations of journal and large-language model platform with complete miss**. The forest plot shows that journal and large-language model (LLM) platform were independently associated with complete miss rate in multivariable regression analysis. Odds ratios with 95% confidence intervals (CI) were shown. The counts and proportions (%) of complete miss by journal and LLM platform are listed on the left side, respectively. *BMJ, British Medical Journal; JAMA, Journal of the American Medical Association; NEJM, The New England Journal of Medicine.*

Journal and LLM platform were associated with complete miss rate in univariable regression, while publication year not. In our multivariable regression analyses journal and LLM platform were both linked to complete miss (**Figure 3**). *NEJM* was associated higher complete miss rates than *BMJ* (p<0.001), but other journals were not. Compared with ChatGPT, Grok was associated a lower compete miss rate (better performance, p<0.001) but others associated with higher rates (worse performance, p<0.001 for all).

**Individual validity metrics**

Across the 400 references retrieved by each LLM platform (100 per journal), the performances of LLM platforms varied significantly in all individual validity metrics by journal or publication year (**Table 1**), except relevance score.

LLM platforms significantly differed in obtaining valid DOI, PubMed ID and Google Scholar links by journal (p<0.001 for all) and publication year (p=0.001, 0.002 and 0.005), respectively. They also differed in relevance scores by journal (p=0.048), but not publication year (p=0.62).

Among the 4 journals, Gemini retrieved accurate bibliographic data of articles least frequently from *BMJ* (0/100 for both PubMed ID and Google Scholar link) and *NEJM* (4/100 for PubMed ID and 12/100 for Google Scholar link). By contrast, Grok and ChatGPT retrieved accurate data more frequently than other LLMs.

**Discussions**

This quantitative comparative study of LLM-assisted medical literature showed modest overall performance of LLMs and significant variability in retrieval accuracy across platforms and journals. These findings align with prior reports documenting high rates of fabricated or incorrect citations in LLM outputs,[8, 9, 14] and extend them by demonstrating that such issues persist even when prompts explicitly request structured bibliographic metadata and data verification. We also revealed the associations of LLM platform and journal with errors in LLM assisted literature retrieval.

Our findings have several implications for LLM-assisted literature research. First, LLM-assisted reference generation/retrieval should be manually verified, particularly when relying on a single platform. References' metadata are often concerning, such as DOI, PubMed ID and Google Scholar links. The variability of LLM performance, as shown by us and others, [4, 18] also highlights the value of using more than one LLM platform at the time. This practice will not only provide more references but the opportunity of cross-checking.

Second, bibliographic data should always be checked even in the LLM-retrieved references with correct titles and sound relevance. Earlier works reveal issues of retrieving irrelevant or less-relevant references by LLM.[11] Much attention has thus been put on relevance of the retrieved ones. However, we show that contemporary LLMs approximate topical relevance reasonably well but struggle with retrieving bibliographic data. The frequent retrieval of plausible references but incorrect bibliographic data could be misleading, offers false assurance, and reinforces the ethical and practical concerns on unsupervised LLM-retrieved references.[16, 19]

Third, accuracy variability by journal also suggests that authors and journal editors should more carefully write and edit abstracts for better literature retrieval, respectively. Longer abstracts may be considered for their associated more information and better medical literature retrieval. Indeed, *NEJM* has shortest abstracts (250 words), as compared with *BMJ* (300 words), *Lancet* (300 words) and *JAMA* (350 words), and was here linked to lower retrieval accuracy.

Finally, explicit prompts to double check references were used here but appear not effective in improving retrieval accuracy. Therefore, these prompts may be tried with caution. More research is needed to better understand and improve LLM-assisted literature retrieval.

This study has limitations. We evaluated only free versions of LLM platforms and focused on a simple prompt design, which may not capture the full range of model capabilities. LLM capacities also improve quickly and may have reduced its errors as this study is being reviewed. Moreover, our relevance scoring, while systematic and mostly objective (e.g., citation by the original article), involved human judgment and may be subjective and biased. Finally, only abstracts of the original articles were subject to the LLM search but may have limited its performance. Future works on full articles will be interesting, but take much more time and seem beyond scope of this study.

In conclusion, while LLMs hold promises as tools for biomedical literature exploration, their current performance in medical literature search and retrieval is highly variable and frequently unreliable. Until robust hallucination mitigation, verification systems, and trustworthy AI frameworks are widely implemented,[20, 21]  LLMs should be used as assistive, not authoritative, tools in biomedical scholarship.

## Acknowledgments

**Funding**

This work was supported by the National Cancer Institute, National Institutes of Health (R37CA277812 to LZ). The funder of the study had no roles in study design, data collection, data analysis, data interpretation, or writing of the report. The corresponding author had full access to all the data in the study and had final responsibility for the decision to submit for publication.

**Author Contributions Statement**

Study conceptualization and design, ensuring the data access, accuracy and integrity (LZ), and manuscript writing (JG and LZ). All authors contributed to the writing and revision of the article and approved the final publication version.

**Conflicts of Interest**

The authors declare no other conflict of interests.

**Data Availability Statement**

The data is available from the corresponding authors on reasonable request.

**Compliance with ethical standards**

This is not a human study.

**Table 1. Characteristics of LLM platforms' performances in retrieving references relevant to the original articles randomly selected from 4 leading medical journals.**

|  | ChatGPT | Deepseek | Gemini | Grok | Perplexity | Total | P-value |
|---|---|---|---|---|---|---|---|
| **DOI** | | | | | | | |
| Journal | | | | | | | |
| BMJ (n=100) | 62 | 23 | 3 | 73 | 9 | 170 | <0.001 |
| JAMA (n=100) | 49 | 13 | 13 | 72 | 19 | 166 | |
| Lancet (n=100) | 35 | 23 | 30 | 53 | 22 | 163 | |
| NEJM (n=100) | 47 | 11 | 7 | 55 | 5 | 125 | |
| Year | | | | | | | |
| 2024 (n=200) | 118 | 26 | 25 | 111 | 32 | 312 | 0.001 |
| 2025 (n=200) | 75 | 44 | 28 | 142 | 23 | 312 | |
| **PubMed ID** | | | | | | | |
| Journal | | | | | | | |
| BMJ (n=100) | 47 | 13 | 0 | 50 | 5 | 115 | <0.001 |
| JAMA (n=100) | 25 | 4 | 4 | 29 | 12 | 74 | |
| Lancet (n=100) | 33 | 10 | 19 | 34 | 10 | 106 | |
| NEJM (n=100) | 26 | 6 | 4 | 33 | 8 | 77 | |
| Year | | | | | | | |
| 2024 (n=200) | 75 | 9 | 16 | 66 | 24 | 190 | 0.002 |
| 2025 (n=200) | 56 | 24 | 11 | 80 | 11 | 182 | |
| **Google Scholar link** | | | | | | | |
| Journal | | | | | | | |
| BMJ (n=100) | 78 | 69 | 0 | 88 | 13 | 248 | <0.001 |
| JAMA (n=100) | 66 | 29 | 4 | 92 | 25 | 216 | |
| Lancet (n=100) | 16 | 43 | 40 | 86 | 5 | 190 | |
| NEJM (n=100) | 60 | 36 | 12 | 80 | 4 | 192 | |
| Year | | | | | | | |
| 2024 (n=200) | 115 | 85 | 39 | 166 | 16 | 421 | 0.005 |
| 2025 (n=200) | 105 | 92 | 17 | 180 | 31 | 425 | |
| **Relevance Score** | | | | | | | |
| Journal | | | | | | | |
| BMJ (n=100) | 32 | 17 | 4 | 50 | 9 | 112 | 0.048 |
| JAMA (n=100) | 34 | 14 | 15 | 41 | 20 | 124 | |
| Lancet (n=100) | 21 | 6 | 12 | 25 | 18 | 82 | |
| NEJM (n=100) | 31 | 12 | 14 | 33 | 14 | 104 | |

| Year | | | | | | | |
|---|---|---|---|---|---|---|---|
| 2024 (n=200) | 60 | 26 | 30 | 80 | 32 | 228 | 0.62 |
| 2025 (n=200) | 58 | 23 | 15 | 69 | 29 | 194 | |

Each number represents the count of valid or relevant references for the performance metric. DOI, Digital object identifier; *JAMA, Journal of the American Medical Association; NEJM, The New England Journal of Medicine; BMJ, British Medical Journal.* P values were generated using Chi-square or Fisher exact test as appropriate.